%% file: main.tex
\documentclass{article}
\usepackage{spconf,amsmath,graphicx}
\usepackage[usenames,dvipsnames]{color}
\usepackage[hidelinks]{hyperref}

\usepackage{xcolor}         

\title{Detection of manatee vocalisations using the Audio Spectrogram Transformer}
\name{
Stefano Schiappacasse$^{1}$ ,
Taco de Wolff$^{2}$,  
Yann Henaut$^{3}$,  
Regina Cervera$^{4}$,  
Aviva Charles$^{5}$ \& 
Felipe Tobar$^{1}$
\thanks{The dataset used in this work was taken at ZooParc de Beauval \& Beauval Nature in France and provided by El Colegio de la Frontera Sur
(ECOSUR/CONAHCYT), Mexico. This project was funded by Google.org, and the following ANID-Chile grants: Fondecyt Regular 1210606, Advanced Center for Electrical and Electronic Engineering
(Basal FB0008) and Center for Mathematical Modeling (Basal FB210005).}
}
\address{$^{1}$Initiative for Data \& Artificial Intelligence, Universidad de Chile, $^{2}$DeWolff AI, \\$^{3}$Departamento de Conservación de la Biodiversidad, El Colegio de la Frontera, $^{4}$C Minds,\\$^{5}$Centre d’Ecologie et des Sciences de la Conservation}

\begin{document}
\ninept
\maketitle
\begin{abstract}
    The Antillean manatee (\emph{Trichechus manatus}) is an endangered herbivorous aquatic mammal whose role as an ecological balancer and umbrella species underscores the importance of its conservation. An innovative approach to monitor manatee populations is passive acoustic monitoring (PAM), where vocalisations are extracted from submarine audio. We propose a novel end-to-end approach to detect manatee vocalisations building on the Audio Spectrogram Transformer (AST). In a transfer learning spirit, we fine-tune AST to detect manatee calls by redesigning its filterbanks and adapting a real-world dataset containing partial positive labels. Our experimental evaluation reveals the two key features of the proposed model: i) it performs on par with the state of the art without requiring hand-tuned denoising or detection stages, and ii) it can successfully identify missed vocalisations in the training dataset, thus reducing the workload of expert bioacoustic labellers. This work is a preliminary relevant step to develop novel, user-friendly tools for the conservation of the different species of manatees.
\end{abstract}
%

\input{introduction}

\input{background}

\input{methodology}

\input{experiments_results}
\input{conclusions_futurework}

\bibliographystyle{IEEEbib}
\bibliography{references}

\end{document}

%% file: introduction.tex
\section{Introduction}
\label{sec:intro}

\subsection{The conservation of the  manatee}

Manatees are herbivorous marine mammals whose populations are declining dangerously. This is the case for the four currently recognised species: the West African manatee (\emph{Trichechus senegalensis})~\cite{vianna2006}, the Amazonian manatee, (\emph{Trichechus inunguis}), the dwarf manatee (\emph{Trichechus pygmaeus})~\cite{vianna2006, rosas1994, roosmalen2015}, and finally the West Indian manatee (\emph{Trichechus manatus}) which comprises two subspecies: the Florida manatee (\emph{Trichechus manatus latirostris}), and the Antillean manatee (\emph{Trichechus manatus manatus})~\cite{lefebvre2001}. The African and Amazonian species live mainly in rivers with turbid waters, while the West Indian manatee is distributed in variable environments, including coastal areas as well as rivers and swamps~\cite{henaut2022}. This diversity of environments makes it difficult to detect their presence and assess their population, which is essential for conservation programs. This is even more evident for species and populations living in environments hindering visual detection.

Most current methods used to detect manatees are visual, where counting of manatees occurs from camera traps, boats, aircraft and, more recently, even drones. These methods have a number of disadvantages: the noise of the engines or drones tends to stress the animals and make them flee, reduced visibility due to the opacity of the water or sunlight greatly limits detection, and the prohibitive financial costs~\cite{landeo2021, ramos2018}. It is, therefore, essential to develop and evaluate the effectiveness of new methods that are less invasive for the animals and their environment, less costly and better adapted to the environments where manatees live. In this sense, the detection of animals by bioacoustics using animal vocalisations represents an effective tool, particularly for cryptic species, as well as non-invasive and inexpensive that also makes it possible to assess behavioural states and the well-being of animals~\cite{teixeira2019}. An initial study on the African manatee shows that bioacoustics can provide an effective response for the detection and conservation of these mammals~\cite{factheu2023}. Of the different species of manatee currently recognised, the West Indian manatee is the most studied for its vocalisations~\cite{reyes2023, richard2004}. This species is more easily accessible both in the wild~\cite{henaut2022} and in zoos~\cite{henaut2023} and motivates the use of passive acoustic monitoring (PAM) to detect and learn about the behaviour of West Indian manatee via the analysis of the recorded underwater audio.

\subsection{Extracting vocalisations from submarine audio}

Detecting manatee vocalisations is challenging. The presence of environmental noise and vocalisations of other animals in submarine recordings hinder the isolation and identification of manatee calls from general underwater acoustic activity. General methods dedicate significant effort to remove noise from the acquired audio signals, which results in discarding the vast majority of recordings comprising acoustic activity not resembling a vocalisation. 

Classical approaches for detection of manatee vocalisations can be divided into those relying on classical signal processing techniques~\cite{automatic, merchan1} and those leveraging deep learning architectures~\cite{merchan2, rycyk2022}. However, most existing methods follow a standard pipeline comprising a \textit{i) denoising stage}, which removes the noise from the audio, a \textit{ii) detection stage}, which discards audio segments without vocalisations, and a \textit{iii) classification stage}, which decides whether the audio sample features a vocalisation or not. These methodologies require setting up several hyperparameters with direct physical meaning, either from a signal processing or bio-acoustic standpoint. Therefore, expert knowledge and extensive experimentation are necessary to implement the three-stage pipeline appropriately. Critically, this results in the designed models being difficult to implement and thus to reproduce by others, and possibly limited only to scenarios that strictly follow the statistical properties of the training data.

\subsection{Scope and contribution} 

Our approach to acoustic-based manatee detection aims to surmount previous challenges related to implementation and hand-crafting requirements as described above. Furthermore, our aim is to develop a mechanism that not only detects vocalisations but also assists in the detection and labelling made by experts. As we will see in the experimental section, the available dataset is quite conservative when it comes to marking vocalisations in the sense that many noisy or murky vocalisations are left unlabelled. We expect the proposed methodology to identify those candidate vocalisations, originally undetected by the expert, and return them for re-labelling.

To achieve these goals, we implement a \textit{transformer architecture}~\cite{attentionIsAllYouNeed}, a model known to have an enhanced representation capability compared to previous audio analysis frameworks. Specifically, our proposal is based on the Audio Spectrogram Transformer (AST) \cite{ast} and is radically different from those in the literature on manatee vocalisations: in our proposal, the denoising, detection and classification stages are contained in a single model fully trained on a real-world dataset recorded at the \textit{ZooParc de Beauval \& Beauval Nature} in France and provided by \emph{El Colegio de la Frontera Sur} (ECOSUR/CONAHCYT), Mexico. 

Our main contributions are: 
\begin{itemize}
    \item An adaptation of the AST model to handle underwater audio, particularly manatee vocalisations. This implies redesigning the temporal and frequency resolution of the filterbanks. 
    \item Fine-tuning the AST model on a dataset of real-world submarine audio with an extremely unbalanced percentage of positive samples of less than 1\%. 
    \item Experimental assessment of the proposed end-to-end method through the lens of the trade-off between precision and recall. 
    \item An identification of previously unlabelled vocalisations returned for relabelling. This alleviates the workload of the expert human annotators and can also used for re-training. 
    \item A discussion about the impact of the proposal in relationship to existing approaches.
    \end{itemize}

%% file: background.tex
\section{BACKGROUND}
\label{sec:back}

\subsection{Related work and our proposal in context}
\label{sec:related_work}
\begin{figure}[h]
\centering
\includegraphics[width=0.43\textwidth]{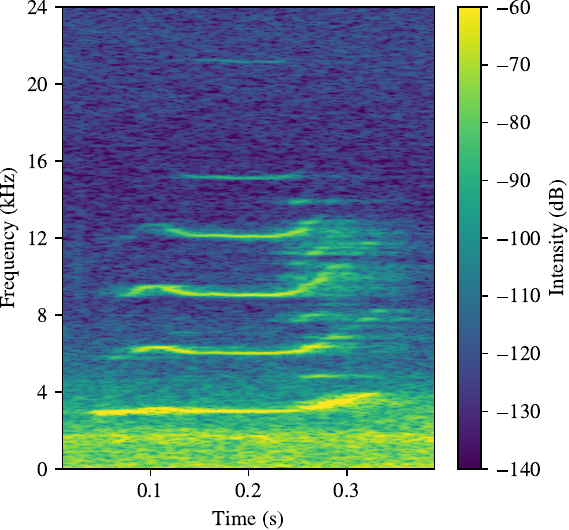}
\caption{Spectrogram of a manatee vocalisation from the Beauval dataset. Typically, manatee calls have strong harmonics reaching ultrasonic frequencies~\cite{ramos2020}, especially for young calves. Each individual has a distinct harmonic pattern and uses different vocalisations for contact calls or mother-calf calls for example.}
\label{label:harmonic_manatee}
\end{figure}

Methodologies for detecting manatee vocalisations from submarine audio rely broadly on signal processing by taking advantage of two main characteristics of manatee vocalisations: the strong harmonic content (see Figure \ref{label:harmonic_manatee}) and the slowly decaying auto-correlation function~\cite{gur2007}. Traditionally, processing raw audio to detect vocalisations operates by first implementing a denoising stage. Then, a detection stage efficiently (in computational terms) discards audio segments that clearly do not contain vocalisations. Lastly, a classification stage identifies manatee vocalisations from non-manatee vocalisations (or non-vocalisations). Next, we present three works relevant to our proposal which follow the above methodology. 

\textbf{Merchan in 2019~\cite{merchan1}}. In the denoising stage, they consider noise and signal subspaces building upon the Karhunen-Loeve transform (KLT) thus performing denoising via projections. Then, their detection stage implements a multi-scale signal analysis to identify harmonic and sub-harmonic components based on the autocorrelation function, passband filters, and duration of the vocalisations. Finally, vocalisations are classified following three criteria: first, the fundamental frequency of vocalisations is summarised via the fast Fourier transform (FFT) peaks, and detection occurs if at least two predefined harmonics are present. Second, it is verified whether the amplitude of the FFT in a given region is below a given threshold, accounting for possible subharmonic components. Third, if only one harmonic is present (as in some manatee vocalisations), the amplitude of the FFT is inspected in the neighbourhood of the peak: if it is also below a specified threshold, detection is positive as indicated by Williams (2005)~\cite{Williams}. The performance reported by this method has an average recall of 77\%~\cite{merchan1}.

\textbf{Castro in 2015~\cite{automatic}}. In their denoising stage, the noisy signal undergoes high-pass filtering at 2 kHz to remove noise outside the vocalisation bandwidth. Subsequently, the Undecimated Discrete Wavelet Transform (UDWT)~\cite{gur2007}, with four decomposition levels, is applied to the signal to then compute the autocorrelation function for the wavelet coefficients at each level. Finally, the Root Mean Square (RMS) is used to distinguish between slowly and quickly decaying envelopes, indicative of manatee calls and noise respectively. Their detection algorithm relies on a matched filter and statistical information concerning the fundamental frequency $F_{0}$ and the peak frequency $F_p$. It splits the denoised signal into windows, where each segment is scored based on its similarity to a manatee call with the score ranging from $-1$ to $1$. A moving average filter is applied to the scores and a fixed threshold detects manatee vocalisations while filtering out sequences of inadequate duration. An average recall of 89\% and a precision of 97\% is obtained.

\textbf{Merchan in 2020~\cite{merchan2}}. Their denoising stage uses Boll’s spectral subtraction method~\cite{boll} to minimise noise---or unwanted artefacts---in signals containing vocalisations. The detection stage is the same as used in~\cite{merchan1}, while their classification stage is based on applying a convolutional neural network (CNN) to an FFT-based spectrogram. The window size was chosen to balance temporal and frequency resolution suitable for the 96 kHz sampling frequency. Regardless of signal duration, zero-padding and centring were applied to ensure uniformity, resulting in spectrograms with a fixed size of 257 × 150 pixels. In this study, the authors tried different architectures for the CNN with varying types of spectrogram generation and obtained an average recall of 87\% with a precision of nearly 95\%.

\textbf{Our approach}. Even though previous works exhibit an appealing performance, they rely on the denoising-detection-classification pipeline. These stages, or at least the first two, involve hyperparameters that are carefully chosen experimentally. Our approach is set apart from previous methods in the sense that we aim for an end-to-end solution where there is no need to remove noise or discard samples. Critically, the proposed model is free from stages to be manually adjusted, but rather all hyperparameters are embedded in the neural architecture and thus learnt from data. Our method is thus expected to be simpler to tune and perform better in terms of generalisation to other unseen datasets---we clarify, however, that assessing generalisation to other cases is beyond the scope of this paper.

\subsection{Processing audio spectrograms using transformers}

\begin{figure}[h]
\centering
\includegraphics[width=0.35\textwidth]{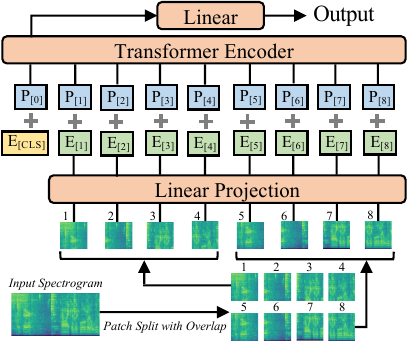}
\caption{Audio Spectrogram Transformer - taken from \cite{ast}.}
\label{fig:ast}
\end{figure}

The core architecture of the proposed method is the Audio Spectrogram Transformer (AST)~\cite{ast}. AST is a Vision Transformer (ViT)~\cite{Vit} adapted to operate on spectrograms rather than general images. Using a time-frequency 2D representation of the recorded audio, the AST leverages the ViT architecture and its parameters, which are pretrained on a large database of general images. The AST thus handles spectrograms as a database of single-channel images which are passed onto the transformer. The architecture of the AST is shown in Figure~\ref{fig:ast}.

AST and ViT follow the same rationale. The input image is divided into fixed-size non-overlapping patches, called \emph{patch embeddings}, where each patch is linearly transformed onto a lower-dimensional vector. This process converts the 2D image into a sequence of embeddings similar to how words are represented in NLP tasks. Also, a \emph{positional encoding} is added to the embedding since the transformer architecture does not inherently account for spatial relationships among tokens. Positional encoding provides information about the position of each patch in the original image. A \emph{transformer encoder}, with its self-attention mechanisms, is then used to capture global dependencies between different patches, allowing each patch to attend to all other patches and to capture long-range dependencies. The output of the transformer encoder is used for classification. A simple fully connected linear layer is attached to predict the class labels.


%% file: methodology.tex
\section{METHODOLOGY}
\label{sec:method}

\begin{figure*}[h]
    \centering
    \includegraphics[width=0.75\linewidth]{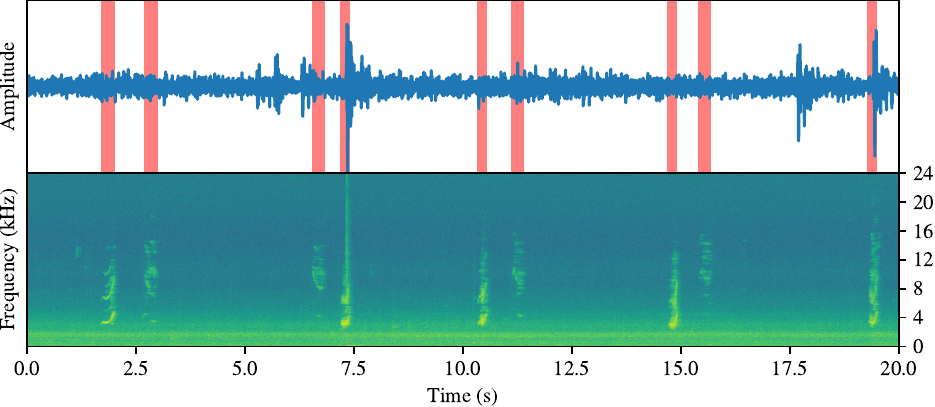}
    \caption{The first 20 seconds of session 9 of the Beauval dataset used for training. The waveform (top) shows the amplitude of the audio file and the manatee vocalisations in red as annotated by an expert bioacoustic. The spectrogram (bottom) in decibels of the same interval shows the high-pitch manatee vocalisations clearly above the background noise.}
    \label{fig_sample}
\end{figure*}

\subsection{Building the classification dataset}
\label{sec:dataset}
Our case study considers a real-world dataset consisting of submarine audio recorded at the \textit{ZooParc de Beauval \& Beauval Nature} in France provided by \emph{El Colegio de la Frontera Sur} (ECOSUR/CONAHCYT), Mexico. The recordings were collected over a three-week period from November 2020 to January 2021 and consist of 20 sessions of roughly 10-minute-long audio, totalling approximately 3 hours sampled at  48 kHz.

Each audio clip has annotations indicating the starting and ending time for the manatee vocalisations detected by a (human) bioacoustic expert. With these labels, each recording contained between 3 to 52 manatee vocalisations, which amounts to less than 1\% of the data, thus making the dataset dramatically unbalanced. Manatee vocalisations are, on average, 240 ms long and typically range from 100 ms to 600 ms. Furthermore, the audio recordings were obtained using an omni-directional hydrophone (Aquarian Audio, H2A-XLR, sensitivity of -180dB re: 1V/$\mu$Pa, frequency range response 20 Hz to 100 kHz) with a Zoom H5 recorder (24-bit quantisation and 48 kHz sampling rate; recording level was set manually to 80). 

A challenge related to this dataset is that, in a learning setting, the provided annotations do not constitute labels that can be used to properly train a classifier. This is because they are only \emph{positive labels}, meaning that unlabelled regions may or may not contain vocalisations. Figure~\ref{fig_sample} shows an example of the recordings in the dataset indicating the audio wave, the spectrogram and regions labelled as vocalisations.

\subsection{Audio preprocessing}

We then focused on producing training and validation sets from the available recordings. To this end, the recordings were split into one-second samples with a 50\% overlap. If one of the annotated manatee calls was present for at least 50\% in the sample, it was labelled as positive, i.e., \emph{containing a vocalisation}. Otherwise, it was labelled as a negative sample. The resulting class labels yielded a highly imbalanced set.  However, this was somewhat alleviated by labelling samples containing a fraction of vocalisation as positive. Furthermore, this approach to labelling is also justified from the perspective of the transformer: it is assumed that the attention module can recognise vocalisations within a larger sample.

Each audio sample was then represented through a filterbank. The frequency and temporal specifications of the bandpass filters, in this case, were chosen so as to appropriately represent the harmonic range of the manatees' vocalisation. Specifically, we adjusted the frame size (with a frame overlap of 50\%) of the filterbanks to obtain samples of size $(64,128)$ using a Hamming window, denoting respectively the number of frequency and time bins. The filterbanks use Mel's scale along the frequency axis, which is a non-linear scale that emphasises relevant frequencies. Experimentally, we confirmed that Mel's frequencies, though calibrated for the frequency range of human voice, performed reasonably well in our experiments. Furthermore, since the fundamental frequency of manatee vocalisations is 2 kHz, with the second harmonic usually containing the highest energy~\cite{chavarria2015}, we dismiss all frequencies below 2 kHz to quell background noise. As our sampling rate is 48 kHz, the frequency range for the filterbanks is thus in the range of [2 kHz, 24 kHz] .


\subsection{Fine-tuning the Audio Spectrogram Transformer}

Our implementation of the AST \cite{ast} is as follows: each filterbank representation of the samples is divided into $N$ 16×16 patches with strides of 10 frames in both time and frequency dimensions, establishing the number of patches and the effective input sequence length for the Transformer (see Fig.~\ref{fig:ast}). Then, each 16×16 patch is flattened into a 1D patch embedding of size 768 through a linear projection layer called the \emph{patch embedding layer}. This linear projection is made with a 2D convolutional layer. As the Transformer architecture lacks the ability to capture input order information, and the patch sequence is not in temporal order, a trainable positional embedding of size 768 is added to each patch embedding. This addition enables the model to grasp the spatial structure of the 2D audio log Mel filterbank. 
Given that AST is tailored to classification tasks, only the Transformer’s encoder is used ~\cite{attentionIsAllYouNeed}. Notably, the original Transformer encoder architecture is adopted without alterations, which aims to facilitate transfer learning in the AST. Specifically, the employed encoder has an embedding dimension of 768, with 12 layers and 12 heads, mirroring those in the Vision Transformer architecture~\cite{Vit}.
Furthermore, an MLP (sigmoid activation) was used to map the classification labels.

The introduced modifications to the AST, which only took part in the time-frequency representation, seek to adapt the transformer to handle the manatee vocalisation data better.

%% file: experiments_results.tex
\section{Experimental validation}
\label{sec:exp}


\subsection{Architecture and computational setup}

We used the AST~\cite{ast} of size \emph{base384} (86.84 million parameters) pretrained on AudioSet~\cite{audioset} and ImageNet~\cite{imagenet}, with a patch splitting stride of 10 for both the frequency and time axes. We used the cross-entropy loss function and the Adam optimiser with a weight decay of $5 \cdot 10^{-7}$, halving the learning rate every 5 epochs starting at $1 \cdot 10^{-6}$, and trained for 25 epochs (approximately 1 hour on an Nvidia GeForce GTX 1080). The number of epochs was chosen to be 25 as we noted a degradation of performance in the testing set afterwards (signs of overfitting). However, it should be noted that training achieved an F1-score of 91\% after 5 epochs already. For reproducibility purposes, our code is available at \href{https://github.com/tdewolff/manatees}{\texttt{github.com/tdewolff/manatees}}.

\subsection{Training considerations and noise injection}

We randomly (uniformly) selected 70\% of the described dataset for training and the remaining 30\% for testing. The dataset was normalised to have zero mean and unit standard deviation. Furthermore, to promote the generalisation capabilities of the model, the training samples were corrupted by random noise with a signal-to-noise ratio of 10 dB to avoid memorisation of the exact training samples. This \emph{noise injection} aims at preventing overfitting and improving out-of-distribution performance. 

To calibrate the precision-recall trade-off as a result of the unbalanced nature of the data, we applied relative weights to each class in the cross-entropy loss function to obtain similar precision and recall results. Specifically, the loss for negative samples is multiplied by $20$ times the ratio $ \frac{\text{\# positive samples}}{\text{\# negative samples}}$. Class weighting could also be employed to improve the recall of the model but to the detriment of the precision. This would be of interest in cases where an expert bioacoustic professional aims to verify only a small subset of possible manatee vocalisations without analysing in detail a large audio recording.

\subsection{Retraining the model with human feedback}

\begin{figure}[t]
\centering
\includegraphics[width=0.4\textwidth]{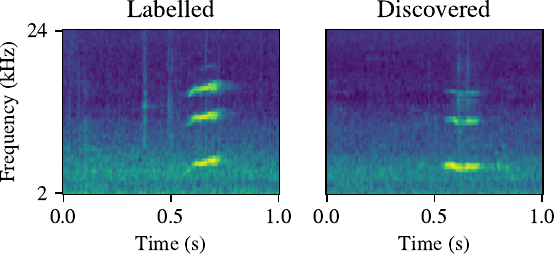}
\caption{Two samples in the training dataset, where only the left one was annotated as vocalisation by the expert. The right sample, however, was determined to be a vocalisation by the proposed model, e.g., a false positive. This way, 357 false positives were reevaluated and subsequently confirmed as a manatee vocalisation by the expert, thus improving the dataset.}
\label{fig:false-positive}
\end{figure}

Recall that the original annotations in the dataset were only positive labels, meaning that samples without annotations could still include vocalisations. Therefore, after training the model with the dataset described in Sec.~\ref{sec:dataset}, a bioacoustic expert revised the false positives found by the model. As confirmed by the domain expert, the vast majority of the false positives were in fact true vocalisations, thus revealing the ability of the AST-based model to identify the vocalisation structure in the data. This way, we dramatically increased the available vocalisations in the dataset from 337 to 694 (from 682 to 1394 training samples). The model was then retrained with the new positive labels, now representing a 6\% of the data, and this feedback loop allowed to critically improve the model's performance, which was initially hindered by the subidentification of high-pitched vocalisations or vocalisations with loud background noise. Fig.~\ref{fig:false-positive} shows two similar samples annotated differently by the original dataset, both identified as vocalisations by the proposed model.

\subsection{Performance assessment}

\begin{table}[t]
    \centering
    \begin{tabular}{l c c c} 
     \hline
     Dataset & Precision & Recall & F1-score \\ 
     \hline
     Original & 0.78 ± 0.03 & 0.76 ± 0.02 & 0.77 ± 0.01 \\
     Human feedback & 0.91 ± 0.01 & 0.94 ± 0.01 & 0.93 ± 0.01 \\ 
     \hline
    \end{tabular}
    \caption{Classification performance of our model on a random selection of 30\% of the original and improved dataset, showing the precision, recall, and their combined F1-score. The model was trained 10 times to show the average and standard deviation of each score. The original dataset corresponds to the results on the initial labelled data by biologists, while the improved dataset corresponds to the results after retraining the model with human feedback on the false positive samples of the original dataset.}
    \label{table:results}
\end{table}

\begin{figure}[t]
\centering
\includegraphics[width=0.35\textwidth]{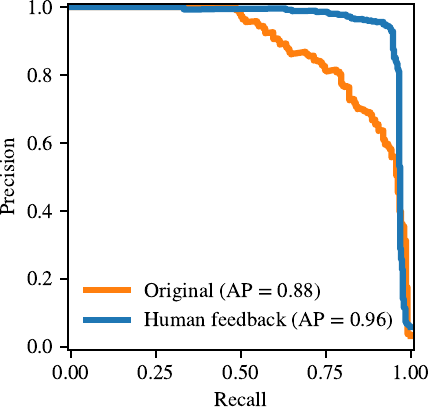}
\caption{Precision-recall relationship for the original (orange) and corrected (blue) datasets as a function of the decision threshold for classification. The average precision (AP), taken over all decision thresholds, is similar to the area under the curve  (higher is better).}
\label{fig:precision-recall}
\end{figure}

Table~\ref{table:results} shows the proposed model's performance in discovering all samples (high recall) and being correct in classifying them (high precision). The classification score is also plotted in Figure~\ref{fig:precision-recall}, which is similar to a Receiver Operating Characteristic (ROC) curve, except that it does not suffer from the class imbalance. This evaluation confirms how incorporating the relabelling (assisted by the proposed model) allows our approach to achieve state-of-the-art performance.

%% file: conclusions_futurework.tex
\section{Discussion}
\label{sec:concl}

\subsection{Novelty and performance }
We have proposed a Transformer-based model for detecting manatee vocalisations from positive-labelled submarine audio recordings. The experimental study validates the performance of the proposed method. It achieved an F1-score of up to 93\% while being trained with a highly unbalanced dataset with only 6\% of positive samples (after preprocessing and relabelling). Though direct comparison against previous studies was not possible due to the consideration of different datasets, our classification performance is similar to that obtained by Merchan et al.~\cite{merchan2} and was usually achieved within 5 epochs (or 15 minutes on a relatively dated Nvidia GTX1080 graphics card). This was made possible by the  AST architecture, pre-trained on ImageNet and AudioSet, which was capable of learning manatee vocalisations within a short training time.

Unlike previous approaches to manatee vocalisation detection, the proposed model does not require preliminary denoising or detection stages described in Sec.~\ref{sec:related_work}. As noted by Castro et al.~\cite{automatic}, standard methods that operate on a denoising-detection-classification pipeline experience a dramatic decrease in performance (from 97\% to 13\%) when denoising is not applied. Therefore, the proposed end-to-end solution is unique in the sense of achieving state-of-the-art performance without requiring prior denoising of the data.

As we discard the denoising and detection stages, our approach aims to discover the entire data-processing pipeline thus setting a more ambitious endeavour than previous methods. As a consequence, this impacts the complexity of the model in terms of the number of parameters. In particular, AST contains more than 86 million parameters, while the CNN used by \cite{merchan1} contains only 4.2 million.

\subsection{Proposed model as an AI assistant to human experts}

Annotating audio files is a tedious task, and files analysed by bioacoustic experts will naturally miss some vocalisations due to noise, the distance between the microphone and the animal,  overlapping of other marine sounds, or simply the presence of other animal acoustic activity. Common causes for failing to classify all vocalisations in an audio file include, e.g., the high-pitch nature of manatee calf vocalisations and loud background noise due to microphone movement. In our experiments, when first training on the original (underlabelled) dataset, our model was able to discover new vocalisations by sifting through a relatively small number of samples classified as false positive; upon verification by the bioacoustic expert, the majority of these samples were found to be in fact manatee vocalisations. 

Integrating a domain expert in the training process, known in the field as \emph{human feedback}, was crucial to improve the dataset's quality and thus the model's performance; specifically from an F1-score of 77\% to 93\%. This validates the proposed model as instrumental to the marine biology community in discovering vocalisations at a reduced workload. Since the audio files used for training were recorded in a zoo and, as a result, include a substantial amount of loud noise~\cite{henaut2023}, it is expected that recordings in the natural habitat of manatees will perform even better.

\subsection{Future work}
Even with the use of a highly imbalanced dataset, our model has shown to learn manatee vocalisations effectively. Increasing the dataset size and diversity will further bolster the model's robustness and accuracy as we move forward. Future research will focus on clustering manatee vocalisations per individual, exploiting the presence of specific frequency components in manatee vocalisations \cite{Williams}, offering more profound insights into their behaviour and social structures. Such detailed data are crucial for estimating population sizes and group dynamics, thereby enhancing conservation efforts, gaining a deeper understanding of the health and dynamics of these key species and their habitats, and supporting the sustainable management of manatee populations in their natural environments.\\
